\documentstyle[preprint,aps]{revtex}
\tightenlines

\begin{document}
\title{Null Singularity Formation by Scalar Fields in Colliding Waves and Black
Holes}
\author{Ozay Gurtug and Mustafa Halilsoy}
\address{Department of Physics, Eastern Mediterranean University G.Magusa, North\\
Cyprus, Mersin 10-Turkey}
\date{\today}
\maketitle

\begin{abstract}
{\small It was believed that when gravitational, electromagnetic and scalar
waves interact, a {\em spacelike} curvature singularity or Cauchy horizon
develops because of mutual focusing. We show with an {\bf exact} solution
that the collision of Einstein-Maxwell-Scalar fields, in contrast to
previous studies, predicts singularities on {\em null} surfaces and that
this is a transition phase between {\em spacelike} singularities and regular
horizons. Divergences of tidal forces in the {\em null} singularities is
shown to be weaker relative to the {\em spacelike} ones. \newline
Using the local isometry between colliding plane waves and black holes, we
show that the inner horizon of Reissner-Nordstrom black hole transforms into
a {\em null } singularity when a particular scalar field is coupled to it.
We also present an analytic {\bf exact} solution, which represents a
Reissner-Nordstrom black hole with scalar hair in between the ergosphere. }
\end{abstract}

\newpage

\section{Introduction}

Colliding plane waves (CPW) provide an excellent test bed toward a better
understanding of singularities in general relativity. Khan and Penrose [1]
considered the collision of two impulsive gravitational waves with parallel
polarizations. They showed that a strong spacelike curvature singularity
develops in the region of interaction. On the other hand, collision of
gravitational waves with ones modified by coupling to sources can yield
totally different results. We cite three examples of such cases. \newline

First, the Bell-Szekeres (BS)[2] solution; this solution represents a
collision of two constant profile electromagnetic (em) shock waves and its
outcome is a quasiregular "singularity" in the interaction region. This is
equivalent to a Cauchy- horizon (CH). \newline

Second, the Chandrasekhar - Xanthopoulos (CX)[3] solution; this solution
describes the collision of two impulsive gravitational waves accompanied by
shock gravitational waves with non-parallel polarizations. It predicts the
development of an event horizon. Analytic extension of the solution across
the horizon reveals the existence of timelike singularities along two
hyperbolic arcs that are locally isometric to the Kerr ergosphere region.
\newline

Third, again as shown by CX [4], coupling em waves to the solution given in
ref. [3], develops a regular null hypersurface which is equivalent to a CH
in the region of interaction. This region is locally isometric to the region
of spacetime in between the two horizons named as event (outer) and Cauchy
(inner) horizons of the Kerr-Newman (KN) black hole (BH).\newline

The CH of BS solution was shown to be unstable against perturbations [5,6].
In ref. [7], it has been shown that, there is a similar inner horizon
instability for BHs and the horizons change to spacelike singularities.
\newline

In brief, these examples conform to the earlier hypothesis that any horizon
formed in CPW is null while any singularity formed is spacelike. \newline

Later Ori [8] found that when the CH of a spinning BH is perturbed the
result is a curvature singularity which has a null character rather than
spacelike one. This new picture, compared to the previously accepted view
attracted many researchers to confirm the same results. Burko [9,10], using
numerical methods confirmed Ori's observation of a regular horizon changing
to a null singularity when he applied a scalar field to a Reissner-Nordstrom
(RN) BH. \newline

The relations between the mathematical theory of BHs and of colliding waves,
motivates us to explore analogous singularities in the space of colliding
waves. \newline

All analyses devoted to singularity formation in the context of CPWs result
in non-null singularities except for the one considered by Ori in plane
symmetric spacetimes [11]. Ori has discussed the null {\em weak} singularity
in plane symmetric spacetimes, his arguments are generalized to CPW only by
employing outgoing perturbation analyses. Although his formulation for plane
symmetric spacetime is justifiable, such an approach is insufficient to find
an exact analytic solution to the Einstein field equations in the region of
interaction. In this sense, the outcome of the outgoing perturbations does
not reflect the real physical situation. The physical reality reveals the
metric component $e^{-U(u,v)}=0 $ on the null hypersurface and causes a
degeneracy in the metric. This degeneracy plays a crucial role on the {\em %
weak} or {\em strong} character of the singularity.\newline

In this paper we close this gap by constructing an example of colliding
Einstein-Maxwell-Scalar (EMS) waves which leads to a null singularity. In
our example of colliding EMS waves leads us to another crucial point. Pure
em shock waves with constant amplitudes yield regular horizons (see BS
solution cited above). On the other hand collision of pure scalar waves as
we show here yields spacelike singularities. We show that by coupling
suitable em fields to the scalar field these space-like singularities are
transformed into null singularities, suggesting that null singularities are
intermediate formations (in other words a transition phase) between horizons
and spacelike singularities. We also show that tidal accelerations at the
null singularities have weaker divergences than in the case of spacelike
singularities and naturally this raises the following question: Is it
possible to manipulate appropriate counter plane waves so that we can
completely eliminate the divergences ?. Although the answer to this question
has so far been negative our example at least verifies that counter- em
terms can be employed to weaken the singularity. \newline

We note that in conformally flat space-times occurence of null singularities
makes some tidal forces to be finite [12]. Our case here is not conformally
flat and all our tidal forces turn out to be divergent, albeit weaker than
before. \newline

The paper is organized as follows. In section II we give the metric for a
new class of colliding EMS and ES fields and their extension to the incoming
regions. In section III we discuss the singularity structure by analysing
the geodesic behaviour and tidal accelerations. In section IV we couple a
scalar field to a linearly polarized version of CX [4] metric that results
in null singularities in the non-spherical extension of the
Reissner-Nordstrom (RN) BH spacetime. We also show a particular scalar hair
confined in the ergosphere of RN BH that does not violate the BH property.
The paper is concluded with a discussion in section V.

\section{A New CPW Geometry with A Null Singularity}

A long time ago Penney gave a solution for EMS fields in spherical symmetry
generalizing the RN solution with the addition of a scalar field [13]. In a
similar manner by replacing spherical symmetry with planar symmetry and
introducing the BS solution instead of RN we obtain a new solution in the
theory of CPWs. We do this by checking all separate Maxwell, scalar and EMS
field equations with appropriate boundary conditions. These include
continuity of metric components with sourceless scalar and em field
equations satisfied at the boundaries. Some first and second derivatives,
however, contain discontinuities (or jumps) as was discussed first by
O'Brien and Synge [14]. Later on BS and CX gave explicit examples of
solutions within the context of Einstein-Maxwell theory of CPW that
justified the discontinuities in some of the energy-momentum components.%
\newline
The inclusion of a scalar field, as we advocate here, is not an exception to
the reality of discontinuities while we cross from the incoming region to
the region of interaction. \newline
Our line element describing the collision of linearly polarized EMS fields
is summarized by,
\begin{equation}
ds^2=\Delta ^{1-A}Z^2 \left(\frac{d\tau^2}{\Delta}-\frac{d\sigma^2}{\delta}
- \delta dx^2\right) -\Delta ^A Z^{-2} dy^2
\end{equation}
where the notation used is,
\begin{eqnarray}
\tau + \sigma &=&2P\sqrt{1-Q^2}  \nonumber \\
\tau - \sigma &=&2Q\sqrt{1-P^2}  \nonumber \\
\Delta &=& 1- \tau^2  \nonumber \\
\delta &=& 1 - \sigma ^2  \nonumber \\
2Z&=& a(1+ \tau )^A + b (1- \tau )^A
\end{eqnarray}
with $P=u \theta (u) $ and $Q = v \theta (v) $, in which $(u,v) $ are null
coordinates and $\theta (x) $ stands for the step function. We choose the
constant $A$, $0<A<1 $ to represent a scalar parameter so that a scalar
charge can be defined by $\sqrt{1-A^2} $. Namely, for $A=0$ we have the
maximum scalar charge of unity, while for $A=1$ the scalar charge vanishes.
The constants $(a,b) $ stand for two additional parameters with $a>0 $ and $%
b>0$. The fact that metric (1) describes colliding EMS fields will be
justified in the sequel. As particular limits, of (1) we observe the
following.\newline

i) For $A=1 $ (and $a=b$ ), it reduces to the well known BS solution of
colliding constant profile em shock waves. This particular solution is known
to possess a horizon in the interaction region. \newline

ii) For $A=0 $, which implies a maximum scalar charge of unity it represents
a collision of Einstein-Scalar (ES) fields that create a spacelike scalar
curvature singularity. This will be discussed separately in section 2.2.
\newline

Our main concern in this paper is to investigate the effect of a scalar
field on the formation of a null singularity. The massless scalar field and
Maxwell equations
\begin{equation}
\partial _{\mu}\left( \sqrt{-g}g^{\mu \nu} \phi _{\nu} \right )=0
\end{equation}
\begin{equation}
\partial_{\mu} \left( \sqrt{-g} F^{\mu \nu} \right)=0
\end{equation}
are both satisfied by the scalar field
\begin{equation}
\phi ( \tau )= \frac{1}{2} \sqrt{1- A^2} \ln\left | \frac{1+ \tau }{1- \tau }%
\right |
\end{equation}
and the em vector potential
\begin{equation}
A_{\mu}=2\delta ^{x} _{\mu} \sqrt{ab} A \sigma ,
\end{equation}
respectively. This implies that for $A=0 $ there exists only a background
scalar field and the solution given in metric (1) represents the collision
of ES fields. A scalar curvature singularity forms in the region of
interaction and has a spacelike character. As we increase $A $ toward unity
the scalar field diminishes and the singularity of the spacetime transforms
to a Cauchy horizon. In the interval $0<A<1 $, we have the case of a null
singularity. \newline
In Appendix A, we have shown that the Weyl and curvature scalars diverge as $%
\tau \rightarrow 1 $ and this is interpreted as a scalar curvature
singularity. However, the fundamental property is that {\em the present
solution does not become singular on any spacelike surface in the region of
interaction}. This can be seen as follows. \newline
We define the singular surface as
\begin{equation}
S( \tau )=1- \tau
\end{equation}
The normal vector to this surface is
\begin{equation}
( \nabla S )^2=g^{\mu \nu}S_{\mu}S_{\nu }= g^{\tau \tau } S^2_{\tau}=\Delta
^A Z^{-2}=(1-\tau ^2)^A Z^{-2}
\end{equation}
As $\tau \rightarrow 1 $ then $( \nabla S )^2 \rightarrow 0 $ which
indicates a null character orthogonal to both of the null directions of the
incoming regions. It is also interesting to check that the line element (1)
becomes null (i.e. $ds^2=0 $ ) as $\tau \rightarrow 1 $ for $0<A<1 $ and $%
0<u,v<1 $. This type of singularity is the first of its kind encountered in
CPWs. In the obtained solution this null singularity emerges as an
intermediate stage between the regular horizon and a space-like singularity.%
\newline
To make this point more clear we employ the following successive
transformations. First we rewrite the metric (1) in terms of new variables $%
\omega $ and $r $ defined by;
\begin{equation}
\omega = \sqrt{ \Delta \delta }=1-u^2-v^2
\end{equation}
\begin{equation}
r=\tau \sigma =u^2-v^2
\end{equation}
and metric (1) becomes
\begin{equation}
ds^2=\Delta ^{1-A}Z^2 \left(\frac{d \omega ^2 - dr^2}{\tau ^2-\sigma ^2} -
\delta dx^2\right) -\Delta ^A Z^{-2} dy^2
\end{equation}
inverting the transformations (9) and (10) leads,
\begin{equation}
2 \sigma ^2 = 1+ r^2 - \omega ^2 -\sqrt{D}
\end{equation}
\begin{equation}
2 \tau ^2 = 1 + r^2 - \omega ^2 + \sqrt{D}
\end{equation}
where $D=(1+r^2-\omega ^2)^2-4r^2 \geq 0 $. \newline
Secondly we set;
\begin{eqnarray}
\omega +r= \xi  \nonumber \\
\omega -r= \eta
\end{eqnarray}
Such that the equations (12) and (13) become,
\begin{equation}
2\sigma ^2= 1- \xi \eta - \sqrt{(1- \xi ^2)(1- \eta ^2)}
\end{equation}
\begin{equation}
2\tau ^2= 1- \xi \eta + \sqrt{(1- \xi ^2)(1- \eta ^2)}
\end{equation}
and the metric (11) takes the following form,
\begin{eqnarray}
ds^2&=& \left[ \frac{1}{2} ( 1+ \xi \eta - \sqrt{(1-\xi ^2) (1- \eta ^2)}) %
\right] ^{1-A} F^2(\xi, \eta) \left[ \frac{d \xi d \eta }{\sqrt{(1-\xi ^2)
(1- \eta ^2)}} \right.  \nonumber \\
& &  \nonumber \\
& & - \left. \frac{1}{2} ( 1+ \xi \eta + \sqrt{(1-\xi ^2) (1- \eta ^2)})
dx^2 \right]  \nonumber \\
& &  \nonumber \\
& & - \left[\frac{1}{2} ( 1+ \xi \eta - \sqrt{(1-\xi ^2) (1- \eta ^2)}) %
\right] ^{A} F^{-2}(\xi, \eta) dy^2
\end{eqnarray}
where $F(\xi,\eta)=\frac{1}{2}Z(\xi,\eta)$. \newline

The corresponding spacetime manifolds with coordinates $(u,v,x,y) $ and $%
(\xi , \eta , x,y) $ are illustrated in Fig.'s 1 and 2 respectively. It
should be noted that, the description of the spacetime in the $(\xi , \eta )$
coordinates breaks down on the null boundaries separating the interaction
region (region IV) from the incoming regions ( regions II and III ), when $%
u=0 $ and $0 \leq v \leq 1 $ or $v=0 $ and $0 \leq u \leq 1 $. These points
correspond to $\xi = 1 $ or $\eta =1 $ respectively and the quantity $\sqrt{%
(1-\xi ^2) (1- \eta ^2)} $ in the metric (17) becomes zero. To avoid this
problem the new variables $\xi $ and $\eta $ are restricted by the following
inequality.
\begin{equation}
0 \leq \xi, \eta < 1
\end{equation}
Physically this means that metric (17) represents the interaction region
only, and hence the null boundaries denoted by the points $A $ and $B $ in
Figure 1. are excluded. \newline
The metric (1) has another interesting property as far as spherical symmetry
is concerned. By taking $Z $ as
\begin{equation}
2Z=a_{0} | 1+ \tau |^{A} - b_{0} |1- \tau |^{A}
\end{equation}
and using the transformations
\begin{equation}
\tau=\frac{m-r}{\sqrt{m^2-Q^2}}, \hspace{0.5cm} x= \phi , \hspace{0.5cm} y=(%
\sqrt{m^2-Q^2}) t , \hspace{0.5cm} \sigma = \cos \theta
\end{equation}
with $Q^2=\frac{ e^2}{A^2} $, where $e$ is an electric charge, transforms
metric (1) into
\begin{equation}
ds^2= e^{-\alpha} dt^2 -e^{\alpha}dr^2 - e^{\beta} d\Omega ^2
\end{equation}
Here we have
\begin{eqnarray}
e^{\alpha}&=&[(r-a_{0})(r-b_{0})]^{-A} \left\{ \frac{b_{0}|r-a_{0}|^A -
a_{0}|r-b_{0}|^A}{b_{0}-a_{0}} \right\}^2  \nonumber \\
& &  \nonumber \\
& & e^{\beta}=[(r-a_{0})(r-b_{0})]e^{\alpha}  \nonumber \\
& &  \nonumber \\
a_{0}&=&m-\sqrt{m^2- \frac{e^2}{A^2}}  \nonumber \\
& &  \nonumber \\
b_{0}&=&m+\sqrt{m^2- \frac{e^2}{A^2}}
\end{eqnarray}
Metric (21) is recognized as the solution by Penney [13], representing the
generalization of RN solution in the presence of a massless scalar field.
For $A=1$ the solution reduces to RN BH.

\subsection{ Extension of the Space-Time into the Incoming Regions}

The metric in region IV (i.e. the interaction region for $u>0, v>0 $) can be
extended across the null boundaries to find the incoming waves that
participate in the collision. For example region II $(u>0, v<0 ) $ is one of
the incoming regions and the metric in this region is given by
\begin{equation}
ds^2= 4(1-u^2)^{\frac{1}{2} - A } Z^2dudv - (1-u^2)\left[ \frac{Z^2}{%
(1-u^2)^{A-1}} dx^2 + \frac{(1-u^2)^{A-1}}{ Z^{2}} dy^2 \right]
\end{equation}
where $2Z=a(1+u)^A +b(1-u)^A$. The non-zero scale invariant Weyl and Ricci
scalars in this region are obtained from those of Appendix A (by imposing $%
v<0$ ) as
\begin{eqnarray}
\Psi^{(0)}_{4}&=& -\frac{A}{2}\left(\frac{a-b}{a+b}\right)\delta (u) - \frac{
\theta (u) }{1-u^2} \left\{\frac{ (2A-1) \left[ 2u^2(1-A) +1\right] }{1-u^2}
\right.  \nonumber \\
& &  \nonumber \\
& & \left. + \frac{A(A-1)}{Z}\left[ a(1-3u)(1+u)^{A-1}+b(1+3u)(1-u)^{A-1} %
\right] \right.  \nonumber \\
& &  \nonumber \\
& & - \left. \frac{ 3A^2(1-u^2) }{2Z^2} \left[a(1+u)^{A-1} -b(1-u)^{A-1} %
\right]^2- \frac{1}{1-u^2}\left[1+2u^2(A-1)\right] \right.  \nonumber \\
& &  \nonumber \\
& & + \left. \frac{A\left[ a(1+u)^{A-1} -b(1-u)^{A-1} \right]}{Z} \right \}
\\
& &  \nonumber \\
\Phi ^{(0)}_{22} &=& \frac{ \theta (u)}{4Z^2(1-u^2)^2} \left\{ (1-A^2) \left[
b^2(1-u)^{2A} + a^2 (1+u)^{2A} \right] \right.  \nonumber \\
& &  \nonumber \\
& & \left. +2ab(1+A^2)(1-u^2)^A \right \}
\end{eqnarray}
The incoming scalar field and the em vector potential are given by
\begin{equation}
\phi(u)=\frac{1}{2}\sqrt{1-A^2}\ln{\left |\frac{1+u}{1-u}\right |},
\end{equation}
and
\begin{equation}
A_{\mu}(u)=2\sqrt{ab}\delta^{x}_{\mu}u,
\end{equation}
respectively. \newline
It is observed that an impulsive gravitational wave component ( i.e. $\delta
(u) $ term), arises only for $a\neq b $ and $A \neq 0 $. For $A=0 $ both the
impulsive term and the em field drops out leaving only a scalar field and
therefore a colliding ES system. It is also clear that an impulsive term
does not exist in the source $\Phi^{(0)}_{22} $ which implies the absence of
a null shell. The positive definiteness of the incoming total energy of our
combined em and scalar fields is crucial and obviously holds true.\newline
The nature of the singularity in the incoming region is investigated by
calculating the Riemann tensors both in local and PPON frames. In local
coordinates the non-zero components are
\begin{eqnarray}
-R_{uxux}&=&e^{V-U} \left [ \Phi^{(0)}_{22} + \Psi^{(0)}_{4} \right ]
\nonumber \\
-R_{uyuy}&=&e^{-V-U} \left [ \Phi^{(0)}_{22} - \Psi^{(0)}_{4} \right ]
\end{eqnarray}
To find the Riemann tensor in a PPON frame, we define the following PPON
frame vectors
\begin{eqnarray}
e^{\mu}_{(0)}&=&\left( \frac{1}{2F}, \frac{1}{2},0,0 \right )  \nonumber \\
& &  \nonumber \\
e^{\mu}_{(1)}&=&\left( \frac{1}{2F}, -\frac{1}{2},0,0 \right )  \nonumber \\
& &  \nonumber \\
e^{\mu}_{(2)}&=& \left( 0,0,-e^{\frac{U-V}{2}},0\right)  \nonumber \\
& &  \nonumber \\
e^{\mu}_{(3)}&=& \left( 0,0,0,-e^{\frac{U+V}{2}}\right)
\end{eqnarray}
Non-zero components in PPON frame that represent the tidal force components
are
\begin{eqnarray}
R_{0202}=R_{0212}=R_{1212}=- \frac{1}{4F^2} \left[\Phi^{(0)}_{22}
+\Psi^{(0)}_{4} \right ]  \nonumber \\
R_{0303}=R_{0313}=R_{1313}=- \frac{1}{4F^2} \left[\Phi^{(0)}_{22} -
\Psi^{(0)}_{4} \right ]
\end{eqnarray}
where
\begin{eqnarray}
e^V&=&(1-u^2)^{1-A}Z^2  \nonumber \\
e^{-U}&=& 1-u^2  \nonumber \\
F&=&(1-u^2)^{1/2-A}Z^2
\end{eqnarray}
It is clear to see that as $u \rightarrow 1 $ all of these components
diverge indicating a coordinate singularity. This is a non-scalar curvature
singularity since all scalars in the incoming region trivially vanish. Note
that as $A \rightarrow 1$ the rate of divergence slows down and when $A =1 $
all the Riemann components become finite.\newline
We also study the geodesics behaviour near the null singular surface. We
choose the Lagrangian $L=(ds/d \lambda)^2 $, where $ds^2$ is the line
element in equation (1) and $\lambda $ is an affine parameter. In addition
to the energy $\epsilon $ we have three conserved momenta. These are
\begin{eqnarray}
P_{x}&=&-e^{V-U} \dot{x} \\
P_{y}&=&-e^{-V-U} \dot{y} \\
P_{v}&=& e^{-M} \dot{u} \\
\epsilon&=&2P_{v} \dot{v} - P^2_{x}e^{U-V} - P^2_{y}e^{U+V}
\end{eqnarray}
where $\epsilon $ may be taken $0 $ for null geodesics and $1$ for timelike
geodesics and dot represents derivative with respect to an affine parameter.
Using equation (34) and (35) we obtain the following equation,
\begin{equation}
2P^2_{v}e^M \frac{dv}{du} = \epsilon +P^2_{x}e^{U-V} + P^2_{y}e^{U+V}
\end{equation}
The geodesic that remains in region II is obtained for $P_{x}=0 $.
Integrating the above equation yields
\begin{equation}
v-v_{0}=\frac{1}{2P^2_{v}} \left ( \epsilon I_{1} + P^2_{y} I_{2} \right )
\end{equation}
where $v_{0}<-\frac{1}{2P^2_{v}} \left ( \epsilon I_{1} + P^2_{y} I_{2}
\right ) $, with
\begin{eqnarray}
I_{1}&=&2a^2B_{\frac{1}{2}} \left[\frac{3}{2}-A,\frac{3}{2}+A \right]+2b^2B_{%
\frac{1}{2}}\left[ \frac{3}{2}+A,\frac{3}{2}-A\right] +\frac{ab \pi}{4}
\nonumber \\
I_{2}&=&\frac{a^4}{2}B_{\frac{1}{2}}\left[\frac{3}{2}-2A,\frac{3}{2}+2A%
\right] +\frac{b^4}{2}B_{\frac{1}{2}}\left[\frac{3}{2}+2A,\frac{3}{2}-2A%
\right]  \nonumber \\
& & + 2a^3bB_{\frac{1}{2}}\left[\frac{3}{2}-A,\frac{3}{2}+A\right] + 2ab^3B_{%
\frac{1}{2}}\left[\frac{3}{2}+A,\frac{3}{2}-A\right]  \nonumber \\
& & + \frac{6a^2b^2 \pi }{32}  \nonumber
\end{eqnarray}
for $0<A<\frac{3}{4} $. Our notation $B_{\lambda} \left[\mu , \nu \right] $
represents the incomplete beta function which is defined in terms of the
hypergeometric function by
\begin{eqnarray}
B_{\lambda} \left[\mu , \nu \right]&=&\int^{\lambda}_{0} t^{\mu -1}
(1-t)^{\nu -1} dt=\mu ^{-1} \lambda ^{\mu} F\left(\mu, 1-\nu;\mu+1;\lambda
\right)  \nonumber \\
& & 0\leq \lambda \leq 1  \nonumber \\
& & \mu , \nu >0
\end{eqnarray}
For $P_{x} \neq 0 $ and $u<1, v $ becomes positive and indicates that
particles starting from $u=0$ in region II can pass to region IV and they
hit the scalar curvature singularity. Null geodesics of region II terminate
their trajectories in the null singularity of the same region. In this
manner the null singular surface does not change the general behaviour of
particles motion in region II. In summary the geodesics behaviour in the
present case is exactly similar to those considered by Matzner and Tipler
[15] for the case of Khan-Penrose and the BS solutions.

\subsection{ The A=0 case and colliding ES waves}

In this section we show explicitly that $A=0 $ in metric (1) describes
colliding ES waves. The generic form of the colliding waves with linear
polarization is described by the line element
\begin{equation}
ds^2=2e^{-M}dudv-e^{-U}\left( e^{V} dx^2 + e^{-V}dy^2\right)
\end{equation}
and the field equations are as follows [16,17].
\begin{eqnarray}
U_{uv}&=&U_{u}U_{v}-2(\Phi ^{(0)}_{11} + 3 \Lambda ^{(0)}) \\
2U_{uu}&=&U^2_{u}+V^2_{u} -2U_{u}M_{u} +4 \Phi^{(0)}_{22} \\
2U_{vv}&=&U^2_{v}+V^2_{v} -2U_{v}M_{v} +4 \Phi^{(0)}_{00} \\
2M_{uv}&=&-U_{u}U_{v} + V_{u}V_{v} + 8\Phi^{(0)}_{11} \\
4\Phi^{(0)}_{02}&=&4\Phi^{(0)}_{20}=2V_{uv}-U_{u}V_{v}- U_{v}V_{u} \\
2\phi_{uv}&=&U_{u}\phi_{v}+U_{v}\phi_{u}
\end{eqnarray}
The solution follows from Eq. (1) upon substitution of $A=0$ which yields
\begin{eqnarray}
e^{-U}&=&1-u^2-v^2=\sqrt{\Delta \delta }  \nonumber \\
e^{-M}&=&\frac{2 \Delta Z^2_{0}}{\sqrt{1-u^2}\sqrt{1-v^2}}  \nonumber \\
e^{V}&=&Z^2_{0}e^{-U}  \nonumber \\
\phi&=&\frac{1}{2}\ln{\left |\frac{1+\tau}{1-\tau}\right |}
\end{eqnarray}
where $2Z_{0}=a+b=const. $ and the null coordinates $(u,v)$ are to be
considered with the step functions $( \theta (u), \theta (v) ) $
respectively. \newline
The non-zero Weyl and Ricci tetrad scalars follow from the Appendix A by
setting $A=0 $,
\begin{eqnarray}
2\Phi^{(0)}_{11}&=&3\Psi^{(0)}_{2}=-6\Lambda ^{(0)}=\phi_{u} \phi_{v}=\frac{%
\theta (u) \theta (v) }{\sqrt{1-u^2} \sqrt{1-v^2} \Delta } \\
\Phi ^{(0)} _{22}&=&\Psi^{(0)}_{4}=\phi ^2_{u}=\frac{ \theta (u) }{%
(1-u^2)\Delta} \\
\Phi ^{(0)} _{00}&=&\Psi^{(0)}_{2}=\phi ^2_{v}=\frac{ \theta (v) }{%
(1-v^2)\Delta} \\
\Phi^{(0)}_{02}&=&\Phi^{(0)}_{20}=0
\end{eqnarray}
It is clear from the Weyl scalars that $\tau = 1$ is a spacetime singularity
and unlike the case of $0<A<1 $ the singular hypersurface is spacelike.
\newline
The incoming components ( for region II) are only $\Phi ^{(0)}_{22}(u) $ and
$\Psi ^{(0)}_{4}(u) $ given by
\begin{equation}
\Phi ^{(0)}_{22}=\Psi^{(0)}_{4}=\frac{ \theta (u) }{(1-u^2)^2}
\end{equation}
Similar components (for region III ) are obtained for $\Phi^{(0)}_{00} $ and
$\Psi^{(0)}_{0} $ by replacing $u \rightarrow v $. We observe that these
components extent into the interaction region from their respective incoming
regions in a continuous manner. However components such as $\Phi^{(0)}_{11} $
and $\Psi^{(0)}_{2} $ arise in the interaction region without counterparts
in the incoming regions. This means there is a discontinuity (or jump) as
far as these components are concerned. This type of discontinuity is
classified as shock-type by CX [18] and is interpreted to originate from the
gravitational shock waves, not from any current sheets or null shells. We
recall in the case of the BS that the Ricci component $\Phi _{02}=a_{0}b_{0}
\theta (u) \theta (v) $, with $a_{0}=const. $ and $b_{0}=const. $, also
suffers from the same discontinuity. For the case of $A \neq 0 $ (and $a
\neq b$ ) we observe from (24) that there is an additional impulsive type of
discontinuity originating from the occurrence of impulsive gravitational
waves.

\section{Geodesics Behaviour and Tidal Accelerations in Region IV.}

As was clarified in the previous section, any test particle that is imported
by one of the incoming waves is forced to enter the region of interaction
and arrives at the singularity in a finite interval of proper time. Now we
shall study the behaviour of test particle geodesics as projected on the $%
(\tau , x,y ) $ subspace. The first integrals of motion from the geodesics
Lagrangian method are,
\begin{eqnarray}
\dot{x}&=&-\frac{P_{x}}{\Delta ^{1-A} Z^2}  \nonumber \\
\dot{y}&=&-\frac{P_{y}}{\Delta^A Z^{-2}}  \nonumber \\
\dot{ \tau }^2&=&\Delta ^A Z^{-2} \delta _{1} + \Delta ^{2A-1} Z^{-4}
P^2_{x} + P^2_{y}
\end{eqnarray}
where $\delta _{1} $ is $0$ (for null) or $1$ (for timelike geodesics) while
$P_{x} $ and $P_{y} $ are constants of motion. ($\Delta $ and $Z $ are given
in equation (2)). Accelerations of the particles are obtained as,
\begin{eqnarray}
\ddot{x}&=& \frac{P_{x}}{\Delta ^{2-A} Z^3}\left \{ a(1+\tau )^A [A-\tau ] -
b (1- \tau )^A [A +\tau ] \right \} \dot{\tau } \\
& &  \nonumber \\
\ddot{y}&=&-\frac{AP_{y}}{2\Delta ^{A+1}} \left \{ a^2(1+ \tau) ^{2A}
-b^2(1-\tau )^{2A} \right \} \dot{\tau } \\
& &  \nonumber \\
\ddot{\tau }&=& -\frac{P^2_{x}}{\Delta ^{2-2A} Z^5 }\left \{ a(1+\tau )^A [A-%
\frac{\tau }{2} ] - b (1- \tau )^A [A +\frac{\tau }{2} ] \right \}  \nonumber
\\
& & - \frac{A \delta_{1}}{2 \Delta ^{1-A} Z^3} \left \{ a(1+\tau )^A - b (1-
\tau )^A \right \}
\end{eqnarray}
in which $\dot{\tau} $ is to be substituted from (52). In order to study the
geodesics behaviour in the vicinity of the null singularity $\tau = 1 $, we
consider expansions in terms of the parameter $\varepsilon = 1- \tau >0$ .
First for the background scalar field case $A=0 $, which makes a spacelike
singularity we have
\begin{eqnarray}
\ddot{\tau} &\sim & \frac{const.}{\varepsilon ^2}  \nonumber \\
\ddot{x} &\sim & \frac{const.}{\varepsilon ^{\frac{5}{2}}}  \nonumber \\
\ddot{y} &\sim & 0
\end{eqnarray}
where the finite terms are denoted shortly by $const. $. The fact that $%
\ddot{y} \sim 0 $ implies that $y$ is proportional to the affine parameter
as $\varepsilon \rightarrow 1 $. A simple analysis reveals also that for $%
A=1 $ all tidal accelerations can be made finite by the choice of suitable
initial conditions. We choose for instance $P_{y}=0 $ within the context of
the BS solution ( i.e. $A=1 $ and $a=b $ ). This originates from the fact
that the norm of the Killing vector associated with the $y$ direction
diverges. For a detailed exposition of these BS geodesics, since it is
beyond our scope here we refer to elsewhere [15,19]. Our main concern here
now is the scalar field in the interval $0<A<1 $. Expansion of the above
accelerations in powers of $\varepsilon $ yield the followings
\begin{eqnarray}
\ddot{\tau} &\sim & \frac{const.}{\varepsilon ^{2-2A}} + \delta _{1} \frac{%
const.}{\varepsilon ^{1-A}}  \nonumber \\
\ddot{x} &\sim & \dot{\tau} \frac{const.}{\varepsilon ^{2-A}}  \nonumber \\
\ddot{y} &\sim & \dot{\tau} \frac{const.}{\varepsilon ^{A+1}}  \nonumber \\
\dot{\tau}&=&\sqrt{P^2_{y}+ c^2_{1}\varepsilon ^{2A-1} }
\end{eqnarray}
in which all $const. $ terms represent finite numbers while the special
constant, $c^2_{1}=\frac{2P^2_{x}}{(a+b)^2} $. We see that as $\varepsilon
\rightarrow 0 $ time-like and null geodesics make no difference. The tidal
accelerations are seen to be worse for $0<A<\frac{1}{2} $ than $\frac{1}{2}%
<A<1$. It turns out as a general rule that as $A$ increases from zero toward
unity the counter scalar field in the collision serves to weaken the
singularity. When it reaches $A=1 $ there is no singularity remaining and $%
\tau =1 $ emerges as a Cauchy horizon.\newline
It is also instructive to calculate the time of fall into the singularity as
measured from the instant of collision. For this purpose we project our line
element into the two dimensional space
\begin{equation}
ds^2=\Delta ^{1-A}Z^2 \left(\frac{d\tau ^2}{\Delta} -d \theta ^2 \right)
\end{equation}
where we assumed $x=const.$,$y=const. $ and $\sigma = \cos \theta $. Now a
geodesics Lagrangian treatment is equivalent to the energy integral
\begin{equation}
\delta_{1}=\Delta ^{1-A} Z^2\left(\frac{\dot{\tau}^2}{\Delta} - \dot{\theta}%
^2 \right)
\end{equation}
where $\delta _{1} $ is $1 (0)$ for timelike (null) geodesics and $\cdot $
represents derivatives with respect to the affine parameter. We obtain as a
result the proper time of fall into the singularity as
\begin{equation}
t_{0}=\int^{1}_{0} \frac{Z^2}{\sqrt{\delta_{1} \Delta ^A Z^2 + \alpha ^2
\Delta^{2A-1}}} d\tau
\end{equation}
where $\alpha $ is a constant of integration associated with the cyclic
coordinate in (58) . The time for null geodesics is obtained as,
\begin{equation}
t_{0}=\frac{a^2}{\alpha }B_{\frac{1}{2}} \left[\frac{3}{2}-A,\frac{3}{2}+A %
\right]+\frac{b^2}{\alpha } B_{\frac{1}{2}}\left[ \frac{3}{2}+A,\frac{3}{2}-A%
\right] + \frac{ab \pi }{8 \alpha}
\end{equation}
For timelike geodesics, $\delta_{1}=1$ we calculate the shortest time for
test particles which is equivalent to choosing $\alpha = 0 $ in equation
(60). The integration gives,
\begin{equation}
t_{0}=aB_{\frac{1}{2}} \left[1-\frac{A}{2},1+\frac{A}{2} \right]+b B_{\frac{1%
}{2}}\left[ 1+\frac{A}{2},1-\frac{A}{2}\right]
\end{equation}
in which $B_{\lambda} \left[\mu , \nu \right] $ is an incomplete beta
function defined in equation (38).\newline
It is interesting to see that the time of fall is determined by the scalar
charge and the incomplete beta function. We recall that the Euler beta
function was encountered in the scattering problems and S-matrix of field
theory. The present problem is also about scattering but instead of the
complete Euler beta function we have the incomplete beta function.

\section{Null Singular CPWs and Black Holes}

In this section, we shall consider another interesting horizon forming CPW
solution found by CX. This solution is locally isometric to the region in
between the inner and event horizons of the KN BH. The solution is described
by the metric,
\begin{equation}
ds^2=X \left( \frac{d \tau ^2}{ \Delta}-\frac{d \sigma ^2}{ \delta} \right)
- \Delta \delta \frac{X}{Y} dy^2 - \frac{Y}{X} \left( dx - q_{2} dy \right)^2
\end{equation}
where
\begin{eqnarray}
X&=&\frac{1}{\alpha^2}\left[(1-\alpha p \tau)^2+\alpha^2q^2\sigma^2\right]
\nonumber \\
Y&=&1-p^2\tau^2-q^2\sigma^2  \nonumber \\
q_{2}&=&-\frac{q\delta}{p\alpha^2} \frac{1+\alpha^2-2\alpha p \tau}{%
1-p^2\tau^2-q^2\sigma^2}
\end{eqnarray}
in which the constant parameters $\alpha,p$ and $q$ must satisfy
\begin{eqnarray}
0<\alpha\leq1  \nonumber \\
p^2+q^2=1
\end{eqnarray}
This metric admits a CH instead of a space-like curvature singularity at $%
\tau=1 $. In ref[20], we have shown the relations between CPWs and the BH
interiors in detail. Now let us consider a class of linearly polarized
versions of the metric(63) which is isometric to the region in between the
horizons of RN BH. Our aim here is to show the existence of a null
singularity by employing a spherically symmetric scalar field $\phi $ which
satisfies, the massless scalar field equation
\begin{equation}
\partial_{\mu}\left( g^{\mu \nu} \sqrt{g} \phi_{\nu } \right)=0
\end{equation}
or equivalently
\begin{equation}
\left[(1- \tau ^2) \phi_{\tau} \right ]_{\tau}- \left[(1- \sigma ^2)
\phi_{\sigma} \right ]_{\sigma}=0
\end{equation}
By shifting the metric function $X \rightarrow Xe^{-\Gamma} $ we can
generate an EMS solution from the known EM solution. The metric function $%
\Gamma $ arises due to the scalar field $\phi$ (see ref [20] for more
detail). As an EM solution we shall employ the diagonal $(q=0)$ version of
metric (63) which is isometric to the RN BH. In terms of $(\tau , \sigma ) $
the metric function $\Gamma $ is obtained from the integrability conditions
\begin{eqnarray}
(\tau ^2- \sigma ^2) \Gamma_{\tau }&=&2 \Delta \delta \left( \tau \phi^2
_{\tau} + \frac{\tau \delta}{\Delta} \phi^2_{\sigma} -2 \sigma \phi_{\tau}
\phi_{\sigma} \right)  \nonumber \\
& &  \nonumber \\
(\sigma ^2- \tau ^2) \Gamma_{\sigma }&=&2 \Delta \delta \left( \sigma \phi^2
_{\sigma} + \frac{\sigma \Delta}{\delta} \phi^2_{\tau} -2 \tau \phi_{\tau}
\phi_{\sigma} \right)
\end{eqnarray}
A general class of separable solution for the scalar field $\phi$ is given
by
\begin{equation}
\phi(\tau , \sigma )= \sum_{n}\{ a_{n}P_{n}(\tau)P_{n}(\sigma) +
b_{n}Q_{n}(\tau)Q_{n}(\sigma) + c_{n}P_{n}(\tau)Q_{n}(\sigma)+
d_{n}P_{n}(\sigma)Q_{n}(\tau) \}
\end{equation}
Where $P$ and $Q$ are the Legendre functions of the first and second kind,
respectively and $a_{n}, b_{n}, c_{n} $ and $d_{n} $ are arbitrary
constants. \newline
{\bf a)} The choice, $Q_{0}(\sigma)=1, P_{0}(\tau)=\frac{1}{2} \ln{|\frac{%
1+\tau}{1-\tau}|}, a_{n}=b_{n}=d_{n}=0 $ and $c_{0}=k, c_{n}=0 (n \neq 0) $
is equivalent to a simple class of spherically symmetric scalar field,
\begin{equation}
\phi ( \tau )= \frac{k}{2} \ln{\left | \frac{1+ \tau }{1- \tau}\right |}
\end{equation}
with $k=constant $. The new metric that represents the interaction region of
linearly polarized colliding EMS fields can be written as
\begin{equation}
ds^2=X e^{- \Gamma}\left( \frac{d \tau ^2}{ \Delta}-\frac{d \sigma ^2}{
\delta} \right) - \Delta \delta \frac{X}{Y} dy^2 - \frac{Y}{X} dx^2
\end{equation}
where
\begin{equation}
e^{- \Gamma}= \left | \frac{1- \tau ^2 }{ \tau ^2 - \sigma ^2} \right |^{k^2}
\end{equation}
With the addition of this scalar field we can see from the energy momentum
scalar $T^{ \alpha }_{ \alpha } $ and $T_{ \mu \nu } T^{ \mu \nu } $, which
are both divergent that $\tau =1 $ is a singularity. Furthermore, the fact
that as $\tau \rightarrow 1 $ the metric function $g^{\tau \tau }
\rightarrow 0 $ for the case $k^2<1 $ implies a null singularity character.
For $k^2 \geq 1 $, however, it retains the spacelike character which is
standard to CPW.\newline
Using the transformations
\begin{equation}
t=m \alpha x, \hspace{.5cm} y= \phi, \hspace{.5cm} \tau=\frac{m-r}{\sqrt{%
m^2-e^2}}, \hspace{.5cm} \sigma=\cos\theta
\end{equation}
with $m \alpha = \sqrt{m^2 - e^2} $ we obtain
\begin{eqnarray}
ds^2 &=& \left ( 1- \frac{2m}{r} +\frac{e^{2}}{r^{2}} \right )dt^2 - e^{-
\Gamma } \left ( 1- \frac{2m}{r} +\frac{e^{2}}{r^{2}} \right )^{-1}dr^2
\nonumber \\
& &  \nonumber \\
& & - r^2 \left( e^{ - \Gamma } d \theta ^2 + \sin ^2 \theta d \phi ^2
\right)
\end{eqnarray}
Obviously this represents an extension of the RN solution with a minimally
coupled scalar field without spherical symmetry. The singularity structure
is investigated by calculating Ricci and Weyl scalars given in Appendix B.
It is clear that coupling of scalar field destroys the BH property and the
event (outer) and Cauchy (inner) horizons become spacelike singular for $k^2
>1 $. and null singular for $k^2<1 $. \newline

{\bf b)} The choice, $b_{n}=c_{n}=d_{n}=0$ equation (69) becomes
\begin{equation}
\phi(\tau, \sigma )=\sum_{n} a_{n}P_{n}(\tau)P_{n}(\sigma)
\end{equation}
This class of scalar field is not spherically symmetric. Surprizingly it is
regular as the CH is approached ($\tau \rightarrow 1 $). For a particular
case we choose $n=2 $, where the scalar field becomes,
\begin{equation}
\phi(\tau, \sigma)=a_{1}\tau \sigma + \frac{a_{2}}{4}(3 \tau^2-1)(3
\sigma^2-1)
\end{equation}
The metric function $\Gamma $ is obtained as
\begin{eqnarray}
\Gamma &=& a_{1}^2 \tau ^2 + \frac{9}{4} a_{2}^2\tau ^2(1-\frac{\tau^2}{2})
- 6a_{1}a_{2} \tau \sigma \Delta \delta +  \nonumber \\
& &  \nonumber \\
& & \frac{\Delta}{4} \left\{ \frac{9}{2} a_{2}^2 \sigma^2(9\tau^2-1)+
\sigma^2 \left( 4a_{1}^2+9a_{2}^2-45a_{2}^2\tau^2 \right) \right \}
\end{eqnarray}
which is finite as the CH is approached $\Gamma (\tau \rightarrow 1)=a_{1}^2
+ \frac{9}{8}a_{2}^2 $. \newline
The Weyl and curvature scalars are also finite in the limit $\tau
\rightarrow 1 $. This choice of scalar field does not effect the CH in the
region of interaction and therefore the spacetime remains regular. For this
class of colliding EMS fields, the analytic extension of the interaction
region is possible. However this analytic extension does not allow us to
interpret this solution outside the ergosphere in the corresponding
transformed BH spacetime. The reason simply is that, the spacetime is not
asymptotically flat and the related energy of the scalar field becomes
unbounded. \newline
The solution obtained in this way can be treated as a RN BH with a scalar
hair confined in between the event and Cauchy horizons.

\section{Discussion}

It is a known fact that CPWs in general relativity result in the creation of
spacelike curvature singularities and rarely in quasiregular singularities
that are equivalent to CHs. CHs are important as far as the analytic
extension of the resulting spacetime is concerned.\newline
In this paper, we have investigated singularities forming in the space of
colliding EMS waves. This is motivated by the appearance of null
singularities in BH spacetimes bombarded by pulses of scalar fields. The
crucial link is the analogy between the mathematical theories of BHs and of
colliding waves. \newline
In our first example we have constructed a new class of colliding EMS waves
that develops null curvature singularity in the region of interaction. In
the problem under consideration, we have found that the null singularity
emerges as a transition phase between a regular horizon and a spacelike
singularity. Geodesics analysis and scalar curvatures reveal a systematic
weakening of divergence in the case of null singularity. \newline
In the second example, we have used the local isometry between the CPW and
BH spacetimes. This method enables us to couple scalar fields to a CPWs,
where analytic exact solution is possible and then transform the resulting
spacetime into BH spacetimes. With this method, we have coupled two types of
scalar fields. In the first case we have used spherically symmetric scalar
field and observed that the scalar field destroys the BH property and the
inner and outer horizons become null singular for $k^2<1 $ and spacelike
singular for $k^2>1 $. As a second example we couple non-spherical scalar
field and observed that the inner and outer horizons remain regular. This
class of solutions can be interpreted as a RN BH with a scalar hair confined
in the ergosphere without a rotational symmetry. \newline
It is important to compare our results to those of previous analyses
[8,9,10]. In our case the null curvature singularities are {\em strong} in
the sense that all tidal forces becomes unbounded. Ori has analysed the
singularity inside a rotating BH using non-linear perturbation theory. His
analyis concluded with a null {\em weak} singularity. Burko confirmed Ori's
results, using numerical methods when he applied scalar field to a RN BH. We
believe that the {\em strong} character in our case arises due to the {\bf %
exact} solutions. On the other hand, Xanthopoulos [21] has shown the
formation of a singularity on the null surface caused by the collision of
plane gravitational and hydrodynamic waves in perfect fluids with equation
of state $\epsilon = p+k, k=constant.$ without a detailed analysis. \newline
Consequently, our overall impression about the CHs is that they do not have
a unique character. They may turn singular or remain regular with respect to
different perturbing potentials. \newline

\section{{\bf Acknowledgement.}\newline
}

We wish to thank Dr. Andrew Shoom and graduate student I. Sakalli for
helpful discussions and comments. \newline

\section{{\bf Figure Captions. } \newline
}

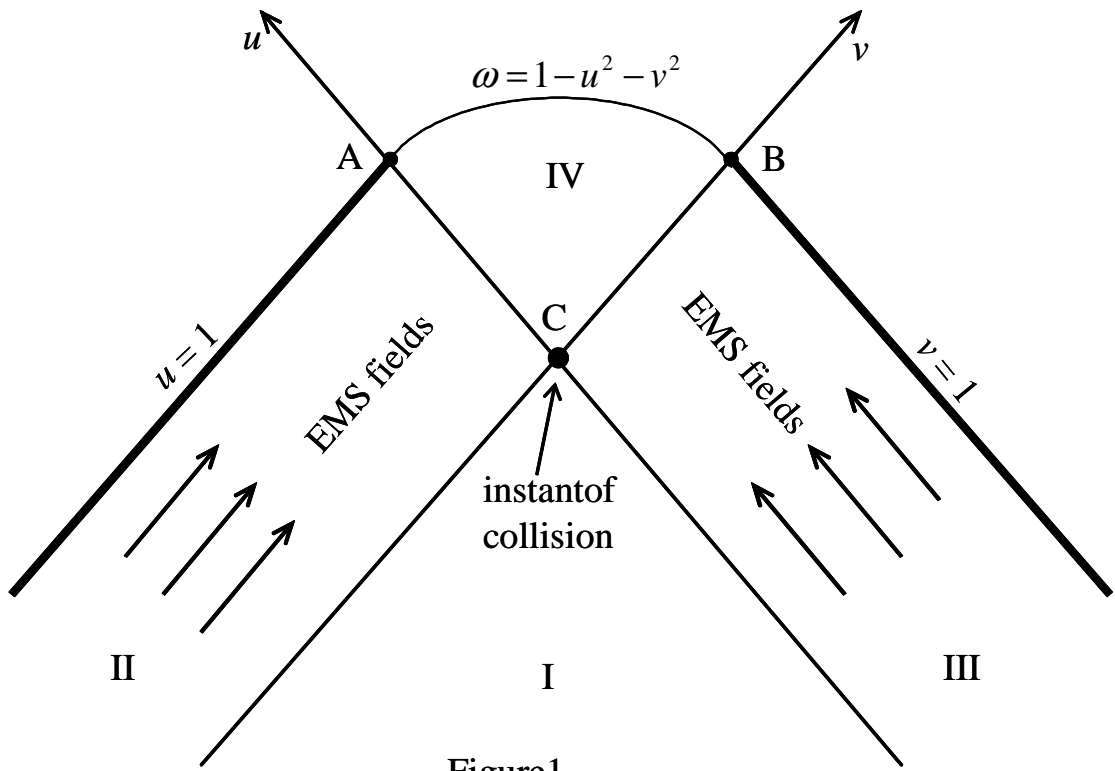
Figure 1. The spacetime diagram for colliding EMS fields. The collision
occurs at point $C $ where $u=v=0 $. In the problem considered the surface $%
\omega = 1-u^2-v^2=0 $ or $\tau = 1 $ represented by the arc $AB $, is null
rather than spacelike. \newline

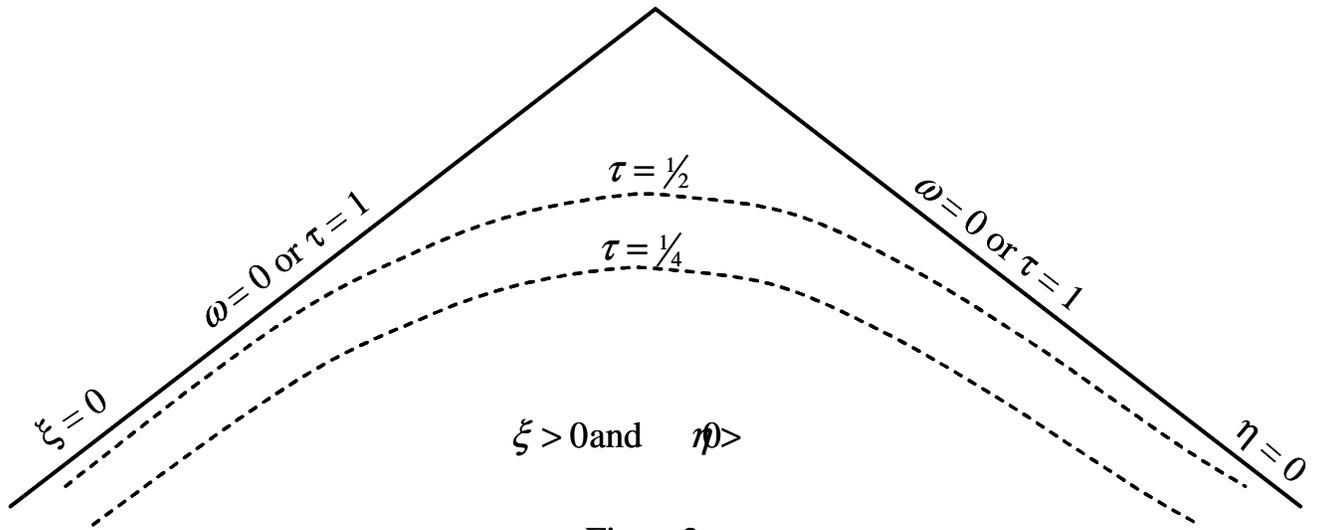
Figure 2. The projection of region IV in $(\xi , \eta ) $ plane. With the
transformation the null character of the arc $AB $ in Figure 1 becomes more
clear. $\xi = \eta = 1 $ is the instant of collision corresponds to $u=v=0 $
or $\tau = 0 $ and $\omega = 1 $. However, $\xi = \eta = 0 $ corresponds to $%
\omega = 0 $ or $\tau = 1 $ where the null curvature singularity occurs.
\newline

\section*{Appendix A: \newline
Properties of the EMS Geometry.}

The non-zero scale invariant Weyl, Ricci and curvature scalars of the
collision of EMS fields are obtained as
\begin{eqnarray}
-2\Psi^{(0)}_{0}&=& 2A \left \{ \frac{u\theta (u)}{\sqrt{1-u^2}} + \frac{%
\sqrt{1-u^2} \left [ a(1+u)^{A-1}-b(1-u)^{A-1} \right ]}{2\left [%
a(1+u)^{A}+b(1-u)^{A}\right ]} \right \} \delta (v)  \nonumber \\
& &  \nonumber \\
& & - \frac{u\theta (u) \theta (v)}{(1-v^2)^{\frac{3}{2}}} \left \{ \frac{%
(2A-1)\tau }{\Delta }-\frac{\sigma }{\delta} + \frac{A \Omega }{Z} \right \}
\nonumber \\
& &  \nonumber \\
& & + \frac{ \theta (v)}{1-v^2} \left \{ \frac{(2A-1)[2\tau ^2(1-A)+1]}{%
\Delta } \right.  \nonumber \\
& &  \nonumber \\
& & + \left. \frac{A(A-1)}{Z}\left [ a(1-3\tau )(1+\tau )^{A-1} + b(1+3\tau
) (1-\tau )^{A-1} \right ] \right.  \nonumber \\
& &  \nonumber \\
& & \left. - \frac{3A^2\Delta }{2Z^2} \Omega ^2 - \frac{1}{\delta} -\frac{v}{%
\sqrt{1-v^2}} \left [ \frac{(2A-1)\tau }{\sqrt{\Delta} }+\frac{\sigma }{%
\sqrt{\delta}} +\frac{A\Omega \sqrt{\Delta}}{Z} \right ] \right \} \\
& &  \nonumber \\
\Psi^{(0)}_{4}&=& \Psi^{(0)}_{0}(u\leftrightarrow v) \\
& &  \nonumber \\
\Psi^{(0)}_{2} &=& \frac{\theta (u) \theta (v) }{\sqrt{1-u^2} \sqrt{1-v^2}}%
\left\{ \frac{1-A}{ \Delta }+\frac{A}{4Z^2} \left \{ a^2(1+ \tau )^{2A-1} +
b^2 (1- \tau )^{2A-1} \right.\right.  \nonumber \\
& &  \nonumber \\
& & \left. \left.+ 2ab (1-2A) \Delta^{A-1} \right \}\right\} \\
& &  \nonumber \\
4 \Phi ^{(0)}_{00}&=& \frac{ \theta ( v) }{(1-v^2) \Delta Z^2 } \left \{
\left ( 1 -A^2\right ) \left [ a^2(1+\tau)^{2A} + b^2(1-\tau )^{2A} \right ]
\right.  \nonumber \\
& &  \nonumber \\
& & \left. + 2ab\left (1 +A^2 \right ) \Delta ^A \right \} \\
& &  \nonumber \\
4 \Phi ^{(0)}_{22}&=& \frac{ \theta ( u) }{(1-u^2) \Delta Z^2 } \left \{
\left ( 1 -A^2\right ) \left [ a^2(1+\tau)^{2A} + b^2(1-\tau )^{2A} \right ]
\right.  \nonumber \\
& &  \nonumber \\
& & \left. + 2ab\left (1 +A^2 \right ) \Delta ^A \right \} \\
& &  \nonumber \\
\Phi^{(0)}_{02}&=& \frac{abA^2\theta (u) \theta (v) \Delta ^{A-1}}{\sqrt{%
1-u^2} \sqrt{1-v^2}Z^2} \\
& &  \nonumber \\
\Phi ^{(0)}_{11}&=& \frac{ (1-A^2) \theta (u) \theta (v)}{2\sqrt{1-u^2}
\sqrt{1-v^2}\Delta } \\
& &  \nonumber \\
\Lambda ^{(0)}&=&\frac{ (A^2-1) \theta (u) \theta (v)}{6\sqrt{1-u^2} \sqrt{%
1-v^2}\Delta }
\end{eqnarray}
where $\Omega = a(1+\tau )^{(A-1)} -b(1- \tau )^ {(A-1)} $. \newline
For $A=1 $ and $a=b$, we recover the collision of Maxwell fields which is
known as the BS solution. At $\tau =1 $ there exist a CH in place of a
curvature singularity. For $A = 0$, we have the geometry that represents the
collision of ES fields which exhibits a spacelike curvature singularity at $%
\tau = 1 $. \newline

For $0<A<1 $ we have the geometry that represents the collision of EMS
fields. Before the collision we have the Weyl scalars $\Psi ^{(0)}_{0} $ and
$\Psi ^{(0)}_{4} $ in region III and II respectively. After the collision we
have $\Psi ^{(0)}_{0} , \Psi ^{(0)}_{4} $ and $\Psi ^{(0)}_{2} $. This
indicates that part of the waves are transmitted in the region of
interaction, part are reflected by each other and part of the incoming waves
transforms into a Coulomb-like $( \Psi ^{(0)}_{2} )$ gravitational waves due
to the non-linear interaction. Some of the Ricci scalars arise
discontinuously in the interaction region. For instance, the energy momentum
component $\Phi ^{(0)}_{11} $ arises in this manner which has no counterpart
before the instant of collision. \newline
The non-zero energy momentum components in terms of sources are,
\begin{eqnarray}
4\pi T_{uu}&=&e^{-M} \Phi _{22}=\Phi^{(0)}_{22}  \nonumber \\
& &  \nonumber \\
4\pi T_{vv}&=&e^{-M} \Phi _{00}=\Phi^{(0)}_{00}  \nonumber \\
& &  \nonumber \\
4 \pi T_{xx}&=& e^{ U-V} \Phi _{02}+ e^{V-U}\left[ \Phi _{11}-3\Lambda
\right ]  \nonumber \\
& &  \nonumber \\
4 \pi T_{yy}&=& -e^{ U+V} \Phi _{02}+ e^{-V-U}\left[ \Phi _{11}-3\Lambda
\right ]
\end{eqnarray}
where
\begin{eqnarray}
e^{-M}&=& \frac{2\Delta ^{(1-A)} Z^2}{\sqrt{1-u^2} \sqrt{1-v^2}}  \nonumber
\\
& &  \nonumber \\
e^V&=&\Delta ^{(\frac{1}{2}-A)} Z^2 \delta ^{\frac{1}{2}}  \nonumber \\
e^{-U}&=& \sqrt{\Delta \delta}  \nonumber
\end{eqnarray}
in which $\Delta, \delta $ and $Z $ is given in equation (2). \newline

\section*{Appendix B: \newline
The Weyl and Maxwell Scalars.}

The nonzero Weyl and Maxwell scalars of the RN BH coupled with a spherically
symmetric scalar field given in equation(70) are calculated and found as
follows.
\begin{eqnarray}
\Psi_{1}&=& -\Psi_{3}= \frac{k^2(m^2-e^2)(mr-e^2)\cos \theta \sin \theta
e^{\Gamma}}{2r^3\sqrt{(r-m)^2-(m^2-e^2)}[(r-m)^2-(m^2-e^2)\cos^2 \theta]}
\nonumber \\
& &  \nonumber \\
\Psi_{0}&=&\Psi_{4}=\frac{k^2(m^2-e^2)(r-m)\sin ^2 \theta
[m(r-m)+m^2-e^2]e^{ \Gamma}}{2r^3[(r-m)^2-(m^2-e^2)][(r-m)^2-(m^2-e^2)\cos^2
\theta]}  \nonumber \\
& &  \nonumber \\
\Psi_{2}&=&\frac{k^2(m^2-e^2) e^{\Gamma}}{[(r-m)^2-(m^2-e^2)]} \left \{
\frac{1}{3r^2} - \frac{(mr-e^2)(r-m)\sin ^2 \theta }{2r^3[(r-m)^2-(m^2-e^2)%
\cos^2 \theta]} \right \}  \nonumber \\
& &  \nonumber \\
& & -\frac{(mr-e^2)e^{\Gamma}}{r^4}  \nonumber \\
& &  \nonumber \\
\Phi_{00}&=&\Phi_{22}=\frac{k^2(m^2-e^2)e^{\Gamma}}{2r^3[(r-m)^2-(m^2-e^2)]}
\nonumber \\
& &  \nonumber \\
\Phi_{11}&=& \frac{e^{\Gamma}}{2r^2} \left \{ \frac{e^2}{r^2} - \frac{%
k^2(m^2-e^2)}{2[(r-m)^2-(m^2-e^2)]}\right \}  \nonumber \\
& &  \nonumber \\
\Lambda&=& \frac{k^2(m^2-e^2)e^{\Gamma}}{12r^2[(r-m)^2-(m^2-e^2)]}
\end{eqnarray}
where
\begin{equation}
e^{\Gamma }=\left | \frac{(r-m)^2-(m^2-e^2)\cos^2 \theta}{[(r-m)^2-(m^2-e^2)]%
} \right |
\end{equation}

\end{document}